\documentclass[final,5p,times]{elsarticle}

\usepackage{amssymb}
\usepackage{amsmath}
\usepackage{xcolor}
\usepackage{graphicx}% Include figure files
\usepackage{dcolumn}% Align table columns on decimal point
\usepackage{bm}% bold math
\usepackage{nicefrac}

\usepackage[hidelinks]{hyperref}
\usepackage{mathtools}
\DeclarePairedDelimiter\bra{\langle}{\rvert}
\DeclarePairedDelimiter\ket{\lvert}{\rangle}
\DeclarePairedDelimiterX\braket[2]{\langle}{\rangle}{#1 \delimsize\vert #2}

\usepackage{bbold}

\usepackage{amsmath}
\DeclareMathOperator{\Tr}{Tr}

\usepackage{ifthen}

\newcommand{\braW}[2]{\bra{\mathbf{#1}#2}}
\newcommand{\ketW}[2]{\ket{\mathbf{#1}#2}}

\newcommand{\braU}[3][]{ \ifthenelse {\equal{#1} {} }
                         {  \bra{u_{#2\mathbf{#3}}} }
                         {  \bra{u^{\mathrm{#1}}_{#2\mathbf{#3}}} }
                       }

\newcommand{\ketU}[3][]{ \ifthenelse {\equal{#1} {} }
                         {  \ket{u_{#2\mathbf{#3}}} }
                         {  \ket{u^{\mathrm{#1}}_{#2\mathbf{#3}}} }
                       }

\newcommand{\expi}[1]{ \mathrm{e}^{\mathrm{i} #1} }
\newcommand{\exmi}[1]{ \mathrm{e}^{-\mathrm{i} #1} }

\newcommand{\bk}{\mathbf{k}}
\newcommand{\bj}{\mathbf{j}}

\newcommand{\bA}{\mathbf{A}}
\newcommand{\Nbq}{N_{\mathbf{q}}}

\newcommand{\partt}{\partial_t}
\newcommand{\rfix}{\rho^{\mathbf{k}}_{\mathrm{S}}}
\newcommand{\rmove}{\rho^{\mathbf{k}}_{\mathrm{C}}}

\newcommand{\intBZdk}{\int\limits_{\mathrm{BZ}} \mathrm{d}\mathbf{k}}
\newcommand{\intBZdkq}{\int\limits_{\mathrm{BZ}} \mathrm{d}\mathbf{k}\mathrm{d}\mathbf{q}}
\newcommand{\rhoHk}{\rho^{\mathrm{H}\mathbf{k}}}
\newcommand{\rhoWk}{\rho^{\mathrm{W}\mathbf{k}}}
\newcommand{\Nkc}{N^{\mathbf{k}}_3}
\newcommand{\agap}[0]{A-$\Gamma$-A path}
\newcommand{\pulsewidth}{\tau}

\journal{Computer Physics Communications}

\begin{document}

\begin{frontmatter}

\title{Semiconductor Bloch equations in Wannier gauge with well-behaved dephasing}

\author[inst1]{Martin Th\"ummler\fnref{equal}}
\author[inst2]{Thomas Lettau\fnref{equal}}
\author[inst1]{Alexander Croy}
\author[inst2,inst3]{Ulf Peschel}
\author[inst1,inst4]{Stefanie Gr\"afe}

\affiliation[inst1]{
    organization={Institute of Physical Chemistry, University of Jena},
    postcode={07743},
    country={Germany},
}
\affiliation[inst2]{organization={Institute of Condensed Matter Theory and Optics, University of Jena},
    postcode={07743},
    country={Germany}
}
\affiliation[inst3]{organization={Abbe School of Photonics, University of Jena},
    postcode={07745},
    country={Germany}
}
\affiliation[inst4]{organization={Fraunhofer Institute for Applied Optics and Precision Engineering},
    city={Jena},
    postcode={07745},
    country={Germany}
}
\fntext[equal]{These authors contributed equally to this work.}

% \date{\today}

\begin{abstract}
    The semiconductor Bloch equations (SBEs) with a dephasing operator for the microscopic polarizations are a well established approach to simulate high-harmonic spectra in solids.
    We discuss the impact of the dephasing operator on the stability of the numerical integration of the SBEs in the Wannier gauge.
    It is shown that the standard approach to apply dephasing is ill-defined in the presence of band crossings and leads to artifacts in the carrier distribution.
    They are caused by rapid changes of the dephasing operator matrix elements in the Wannier gauge, which render the convergence of the simulation in the stationary basis infeasible.
    In the comoving basis, also called Houston basis, these rapid changes can be resolved, but only at the cost of a largely increased computation time.
    As a remedy, we propose a modification of the dephasing operator with reduced magnitude in energetically close subspaces.
    This approach removes the artifacts in the carrier distribution and significantly speeds up the calculations, while affecting the high-harmonic spectrum only marginally.
    To foster further development, we provide our parallelized source code.
\end{abstract}

%% Keywords

\end{frontmatter}

%\tableofcontents
\section{\label{sec:intro}Introduction}
Since its first observation in 2011 \cite{ghimire_2011}, high-harmonic generation (HHG) in solids has been intensively studied in a wide range of materials \cite{schubert_sub-cycle_2014,you_anisotropic_2017,luu_extreme_2015,yoshikawa_high-harmonic_2017,liu_high-harmonic_2017}.
Its applications include building compact sources of high-frequency radiation \cite{han_high-harmonic_2016}, optical investigation of the band structures \cite{vampa_all-optical_2015}, and probing phonon dynamics \cite{neufeld_probing_2022}.
Theoretically, HHG in solids is described by various methods \cite{yu_high_2019,yue_2022} ranging from analytical approaches like the saddle-point analysis \cite{hawkins_role_2013,vampa_2014} to computationally very expensive real-time time-dependent density functional theory calculations \cite{tancogne-dejean_impact_2017}.
However, to model decoherence effects microscopically, the semiconductor Bloch equations \cite{lindberg_1988, yue_2022} (SBEs) are a common choice.
Proper modeling of dephasing is crucial as it influences carrier dynamics \cite{chen_real-time_2025}, coherent effects \cite{korolev_unveiling_2024, herrmann_nonlinear_2025}, and, as widely discussed in the literature, the high-harmonic spectrum \cite{vampa_2014,floss_ab_2018}.
The origin of ultrashort dephasing times in the femtosecond range, which are required to match the experiments on high-harmonic generation, has sparked many discussions \cite{floss_ab_2018,kilen_propagation_2020,vampa_2014}.
Some more advanced dephasing models proposed a $\bk$-dependent dephasing time \cite{langer_lightwave-driven_2016,wismer_gauge-independent_2018,korolev_unveiling_2024}.
Predominantly, the dephasing operator is based on the field-free states and is not invariant under a gauge transformation of the electric field \cite{lamb_matter-field_1987}, contrary to the coherent part of the SBEs \cite{foldi_gauge_2017}.
For this reason and because the convergence with respect to the amount of included bands is improved \cite{wu_high-harmonic_2015}, it is favorable to couple to the electric field in the length gauge, where a careful description of the transition dipole elements is required \cite{jiang_effect_2017}.
Since the evaluation of the dipole operator involves a derivative with respect to the crystal momentum \cite{blount_formalisms_1962}, it is necessary to construct a smooth phase for the Bloch states, which are obtained from ab-initio calculations featuring random phases. 
Several solutions for this problem have been proposed, such as the parallel transport gauge \cite{wang_ab_2006,yue_2020} or the use of simpler models with analytical expressions for the band structure \cite{vampa_semiclassical_2015,chacon_circular_2020}.
In recent years, the usage of maximally localized Wannier functions has become the common approach to work with arbitrary band structures \cite{marzari_1997, silva_2019}, which also allows for a well-defined calculation of the dipole operator \cite{wang_ab_2006}. 

In the present work, we study the numerical behavior of the SBEs in Wannier gauge employing the stationary basis (SB) \cite{golde_high_2008,floss_ab_2018} and the comoving basis (CB) \cite{vampa_2014,yue_2020}.
The latter is often referred to as the Houston or adiabatic basis \cite{houston_acceleration_1940,krieger_time_1986,wu_high-harmonic_2015,yue_2022}.
In the CB, the equations for individual $\bk$-points decouple, which allows for an efficient parallel calculation.
However, this comes with an increased computational effort to calculate the Hamiltonian and transition dipole elements for each time step.
We show that by utilizing fast Fourier transforms (FFTs) for the interpolation of the matrix elements, the numerical integration effort of the CB is leveraged to be similar to that of the SB.

The inclusion of the dephasing operator will be handled with particular care.
We call the commonly used form \cite{vampa_2014} constant dephasing operator (CDO) and uncover problems it causes when used in the Wannier gauge.
We prove that in the case of degenerate or crossing bands, the CDO is mathematically ill-defined.
Close to avoided crossings, the integration reveals undesired and sharp kinks in the momentum space that appear in the distribution of the excited carriers.
These kinks are an intrinsic property of the CDO and contribute only marginally to the macroscopic physical observables.
The numerical resolution of these kinks in the CB comes with a huge computational cost, which dominates the numerical effort of the SBE simulation. 
In the SB, the calculations are numerically unstable, and it is hardly possible to converge them.
We rectify these problems by proposing and applying a soothed dephasing operator (SDO), which is a function of the energy differences of the corresponding states and reduces the dephasing to zero for degenerated bands.
We show that the SDO does not significantly alter the dynamics of the SBEs compared to the CDO, except for the suppression of the kinks in the carrier distribution.
In particular, the high-harmonic spectrum remains unchanged.

The study is structured as follows.
In the next section, we review the SBEs in Wannier gauge for the SB and the CB.
This is followed by details on the implementation in Sec.~\ref{sec:implementationDetails}.
In Sec.~\ref{sec:dephasingOperator}, we identify the problems that arise for the CDO, and we introduce the SDO.
Afterwards, we demonstrate its significance in Sec.~\ref{sec:numericalResults} via numerical studies.
Finally, we summarize and conclude with Sec.~\ref{sec:sao}.
Throughout the manuscript, atomic units are used if not otherwise indicated.

\section{\label{sec:theory}Theoretical background}
In this section, we recapitulate the derivation of the SBEs in the Wannier gauge \cite{silva_2019}.
We introduce the Bloch functions $\ket{\Psi^{\mathrm{H}}_{\mathbf{k}n}}$ with crystal momentum $\mathbf{k}$ and band index $n$ and denote their periodic part as $\ketU[H] nk = \exmi{\mathbf{k}\mathbf{r}}\ket{\Psi^{\mathrm{H}}_{\mathbf{k}n}}$, which satisfies
\begin{equation}
    H^{\mathrm{H}}_{0,\mathbf{k}} \ketU[H] nk = E^\mathbf{k}_n \ketU[H] nk
    ,
\end{equation}
where $H^{\mathrm{H}}_{0,\mathbf{k}}$ is the field-free Hamiltonian and $E^\mathbf{k}_n$ is the corresponding eigenvalue. 
The superscript '$\mathrm{H}$' denotes the so-called Hamiltonian gauge.

\subsection{Wannier functions}
The Wannier functions (WF) are calculated based on the ab-initio Bloch functions and are used to interpolate operators from the coarse $\mathbf{q}$-grid to an arbitrary $\bk$-point, e.g., the Hamiltonian \cite{marzari_1997,pizzi_2020}.
The transformation of $\ketU[H] nq$ from the Hamiltonian gauge into the smooth Wannier gauge (denoted by the superscript '$\mathrm{W}$') is described by
\begin{equation}
    \label{eq:WannierTransformation}
    \ketU[W] mq = \sum\limits_{n} U^{\mathbf{q}}_{mn} \ketU[H] nq
,
\end{equation}
where $U^{\mathbf{q}}$ are semiunitary transformation matrices \cite{souza_2001}.
We denote the number of WFs by $n_{\mathrm{W}}$ which is equal to the number of included bands.
The $n$-th WF itself, located at unit cell $\mathbf{R}$, is calculated as
\begin{equation}
    \ketW Rn = \frac{1}{\Nbq} \sum\limits_{\mathbf{q}} \exmi{\mathbf{q}(\mathbf{R}-\hat{\mathbf{r}})} \ketU[W] nq
,
\end{equation}
where the $\mathbf{q}$ sum is carried out over all $\Nbq$ points of the ab-initio Monkhorst-Pack (MP) grid.
They form an orthonormal basis, i.e., $\braket{\mathbf{R}_1 n_1}{\mathbf{R}_2 n_2} = \delta_{\mathbf{R}_1\mathbf{R}_2} \delta_{n_1n_2}$.
Because of the exponential decay of the WFs in position space \cite{kohn_analytic_1959}, one can interpolate the original $\ketU[W] mq$ to an arbitrary point in momentum space $\mathbf{k}$ by taking the inverse transformation
\begin{equation}
    \ketU[W] nk = \sum\limits_{\mathbf{R}} \expi{\mathbf{k}(\mathbf{R}-\hat{\mathbf{r}})} \ketW Rn
    .
\end{equation}
As we aim to express the Hamiltonian including light-matter interaction in the Wannier gauge, we evaluate the matrix elements of the Hamiltonian and the position operator \cite{wang_ab_2006}
\begin{align}
    \label{eq:WanH}
    H^{\mathrm{W}\mathbf{k}}_{mn} &\equiv \braU[W]mk \hat{H}_0 \ketU[W]nk =
        \sum\limits_{\mathbf{R}} \exmi{\mathbf{k}\mathbf{R}} \braW Rm \hat{H}_0 \ketW 0n
    , \\
    \label{eq:WanD}
    \mathbf{D}^{\mathrm{W}\mathbf{k}}_{mn} &\equiv \mathrm{i} \braU[W]mk \nabla_{\mathbf{k}} \ketU[W]nk =
        \sum\limits_{\mathbf{R}} \exmi{\mathbf{k}\mathbf{R}} \braW Rm \hat{\mathbf{r}} \ketW 0n
,
\end{align}
respectively.
For the current operator, we later require the derivative of the Hamiltonian matrix elements
\begin{equation}
    \label{eq:WandHdk}
        \nabla_{\mathbf{k}}H^{\mathrm{W}\mathbf{k}}_{mn} = -\mathrm{i}
        \sum\limits_{\mathbf{R}} \mathbf{R} \exmi{\mathbf{k}\mathbf{R}} \braW Rm \hat{H}_0 \ketW 0n
        ,
\end{equation}
which are evaluated analytically.

\subsection{SBEs in the stationary basis}
We proceed to derive the coherent part of the SBEs in Wannier gauge.
We start with the Hamiltonian in dipole approximation and length gauge
\begin{equation}
    \hat{H}(t) = \hat{H}_0 + \mathbf{E}(t) \cdot \hat{\mathbf{r}}
,
\end{equation}
where $\mathbf{E}(t)$ is the time-dependent electric field.
We express the density matrix via the interpolated Wannier functions as
\begin{equation}
    \hat{\rho} = \intBZdkq\sum\limits_{mn} \rho^{\mathbf{kq}}_{mn} \ket{\Psi^{\mathrm{W}}_{\mathbf{k}n}} \bra{\Psi^{\mathrm{W}}_{\mathbf{q}m}}
    ,
\end{equation}
where we defined $\ket{\Psi^{\mathrm{W}}_{\mathbf{k}n}} = \expi{\mathbf{k}\mathbf{r}} \ketU[W] nk$.
For an arbitrary operator $\hat{O}$, we denote its matrix elements by $O^{\mathbf{kq}}_{mn} = \bra{\Psi^{\mathrm{W}}_{\mathbf{k}m}} \hat{O} \ket{\Psi^{\mathrm{W}}_{\mathbf{q}n}}$.
Employing the von-Neumann equation $\mathrm{i}\partt \hat{\rho} = [\hat{H}, \hat{\rho}]$, we obtain the SBEs for the matrix elements
\begin{align}\label{eq:fixedFrameVonNeumann}
    \mathrm{i}\partt& \rho^{\mathbf{kq}}_{mn}
                                       = \delta_{\mathbf{kq}} \left( [ H^{\mathrm{W}\mathbf{k}} + \mathbf{E}\cdot\mathbf{D}^{\mathrm{W}\mathbf{k}}, \rho^{\mathbf{kk}} ]_{mn}
                                           + \mathrm{i} \mathbf{E}\cdot \nabla_{\mathbf{k}} \rho^{\mathbf{kk}}_{mn} \right)
,
\end{align}
where we used $H_{0, mn}^{\mathbf{kq}}= \delta_{\mathbf{kq}}H^{\mathrm{W}\mathbf{k}}_{mn}$ and $(\hat{\mathbf{r}})^{\mathbf{qk}}_{mn} = (\mathrm{i} \delta_{mn} \nabla_\mathbf{q} + \mathbf{D}^{\mathrm{W}\mathbf{k}}_{mn}) \delta_{\mathbf{qk}}$ \cite{blount_formalisms_1962,yue_2022}.
Here we already dismissed any coupling between different $\mathbf{k}$-points, which would be required if, e.g., the Coulomb interaction is included explicitly \cite{meier_coherent_1994}.
Since the initial density matrix is diagonal in momentum as well as its time derivative, see  Eq.~\eqref{eq:fixedFrameVonNeumann}, we define $\rho^{\mathbf{k}} \equiv \rho^{\mathbf{kk}}$.
The SBEs in the SB reads \cite{silva_2019}
\begin{equation}
    \label{eq:SBE_fixed}
    \mathrm{i}\partt \rho^{\mathbf{k}}
     = [ H^{\mathrm{W}\mathbf{k}} + \mathbf{E}\cdot\mathbf{D}^{\mathrm{W}\mathbf{k}}, \rho^{\mathbf{k}} ]
                                           + \mathrm{i} \mathbf{E}\cdot \nabla_{\mathbf{k}} \rho^{\mathbf{k}}
.
\end{equation}
The microscopic current operator is defined through $\hat{\mathbf{j}} = \mathrm{i} [\hat{\mathbf{r}}, \hat{H}]$, which can be expressed in terms of the Hamiltonian and the position operator to obtain the current as \cite{silva_2019}
\begin{equation}
    \label{eq:SBEfixedCoherent}
    \mathbf{J} = \intBZdk
                    \Tr\left\{\rho^\mathbf{k} \left(\mathrm{i}[\mathbf{D}^{\mathrm{W}\mathbf{k}}, H^{\mathrm{W}\mathbf{k}}] - \nabla_{\mathbf{k}} H^{\mathrm{W}\mathbf{k}} \right) \right\}
.
\end{equation}

\subsection{SBEs in the comoving basis}
We derive the equations of motion for the CB \cite{vampa_2014} from the SB.
To eliminate the gradient of the density matrix in Eq.~\eqref{eq:SBE_fixed}, we first define the transformation of the SB to the CB as
\begin{equation}
    \ket{\Psi^{\mathrm{W}}_{\mathbf{k},n} } 
    \rightarrow 
    \ket{\Psi^{\mathrm{W}}_{\mathbf{k} + \mathbf{A}(t),n} }
\end{equation}
with the vector potential
\begin{equation}
    \mathbf{A}(t) = - \int\limits_{-\infty}^t \mathbf{E}(t')\,\mathrm{d}t'
.
\end{equation}
The matrix elements of the density matrix and their time derivatives transform under this change of basis as
\begin{align}
\rho^{\mathbf{k}} \quad\rightarrow&\quad\rho^{\mathbf{k}+\mathbf{A}},  \nonumber \\
\partial_t \rho^{\mathbf{k}} \quad\rightarrow&\quad \partial_t \rho^{\mathbf{k} + \mathbf{A}} - \mathbf{E} \cdot\nabla_{\mathbf{k}} \rho^{\mathbf{k} + \mathbf{A}} \label{eq:sbcbTransform}
.
\end{align}
We note that in this notation the partial-time derivative does not act on the vector potential anymore.
Inserting Eq.~\eqref{eq:sbcbTransform} into Eq.~\eqref{eq:SBE_fixed} directly leads to the SBEs in the CB \cite{kim_theory_2022}
\begin{equation}
    \label{eq:movingFrameVonNeumann}
    \mathrm{i} \partial_t{\rho}^{\mathbf{k}} = [ H^{\mathrm{W},\mathbf{k} + \mathbf{A}} + \mathbf{E}\cdot\mathbf{D}^{\mathrm{W},\mathbf{k}+\mathbf{A}}, \rho^\mathbf{k} ]
.
\end{equation}
Finally, the current is evaluated as 
\begin{align}
\label{eq:SBEMovingCurrent}
\begin{split}    
    \mathbf{J} = \intBZdk \Tr\big\{ \rho^\mathbf{k} \big(
     % \\[-2ex]
     & \mathrm{i} \big[\mathbf{D}^{\mathrm{W},\mathbf{k}+\mathbf{A}}, H^{\mathrm{W},\mathbf{k}+\mathbf{A}} \big] %\\[-3ex]
     %-& 
     - \nabla_\mathbf{k} H^{\mathrm{W},\mathbf{k}+\mathbf{A}}
     \big)
    \big\}
    .
\end{split}
\end{align}

\subsection{Inclusion of the dephasing operator \label{sec:OrigDephasingOperator} }
Typically, the phenomenological dephasing operator is applied in the Hamiltonian gauge as an additional incoherent term to the time derivative of the density matrix \cite{vampa_2014}
\begin{equation} \label{eq:origDephasing}
    \left( \partial_t\rho^{\mathrm{H}\mathbf{k}}_{mn} \right)_{\mathrm{Deph}} = -  \frac{1 - \delta_{nm}}{T_2} \rho^{\mathrm{H}\mathbf{k}}_{mn}
,
\end{equation}
where $m$ and $n$ are band indices.
To obtain the dephasing operator for the SBEs in Wannier gauge, we first transform the density matrix according to Eq.~\eqref{eq:WannierTransformation} as
\begin{align} \label{eq:rhoWtoH}
    \rhoHk = U^{\mathbf{k}} \rhoWk U^{\mathbf{k}\dagger}
.
\end{align}
The dephasing in the Hamiltonian gauge, calculated via Eq.~\eqref{eq:origDephasing}, is then transformed back to the Wannier gauge and added to the time derivative of the density matrix \cite{silva_2019}.

In their fundamental work, Silva et al. \cite{silva_2019} used a constant time step of 2.5 atomic units.
It was also argued \cite{yue_2020,yue_2022}, that due to necessary basis changes, the computational cost may be reduced by applying the dephasing only after the coherent part was integrated for predefined time interval $\Delta t$.
To this end, the density matrix is updated as
\begin{equation}
  \rho^{\mathrm{H}\mathbf{k}}_{mn} \rightarrow \rho^{\mathrm{H}\mathbf{k}}_{mn} \mathrm{e}^{-\Delta t / T_2},\quad m\neq n
  .
\end{equation}
Being valid for their respective models, these approaches assume that the matrix elements of the dephasing operator vary relatively slowly over the BZ and time.
However, we will show in Sec.~\ref{sec:dephasingOperator} that this assumption does not hold in the vicinity of avoided crossings.
\begin{figure}[ht]
    \centering
    \includegraphics[width=0.99\linewidth]{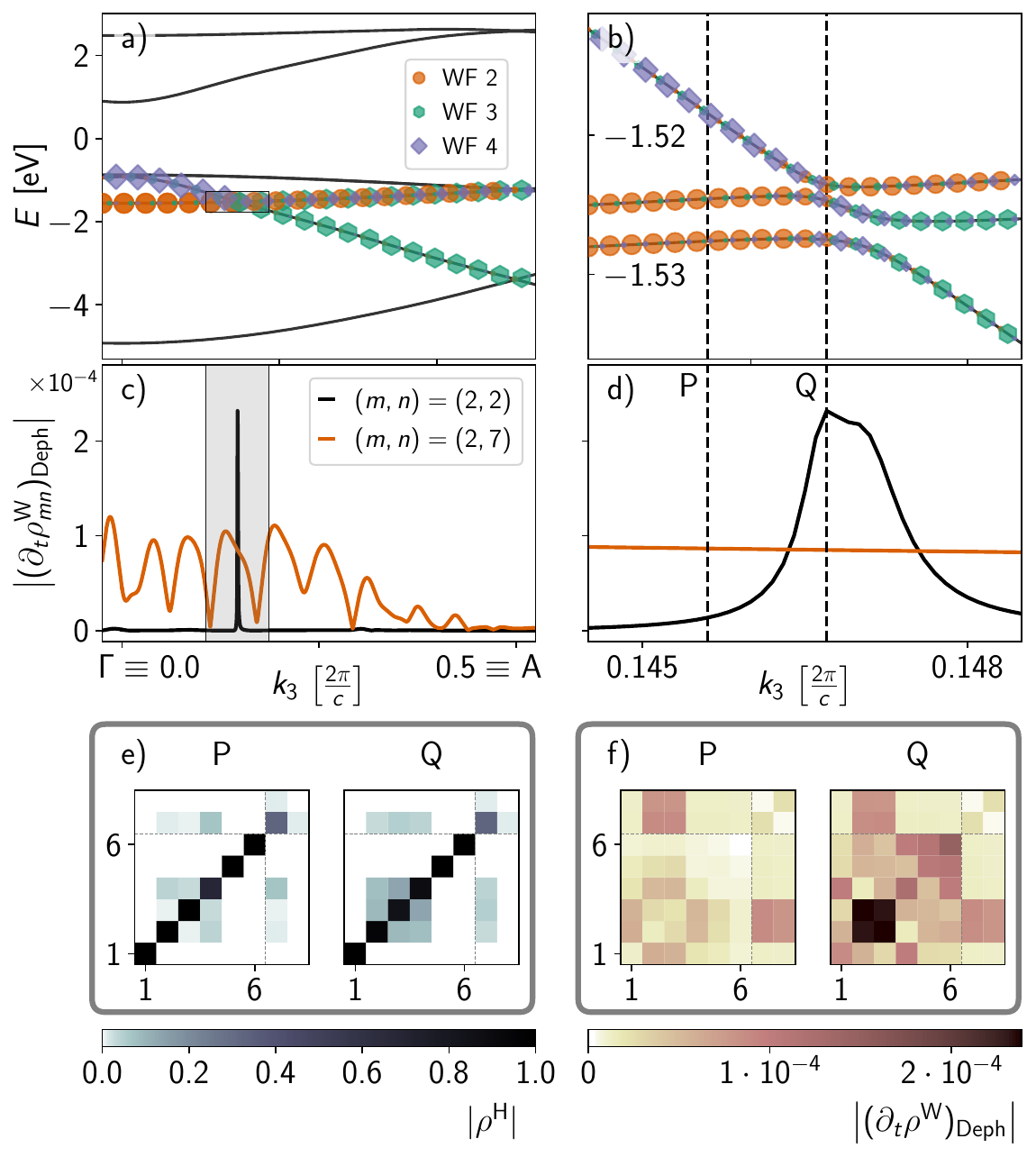}
    \caption{Band structure of CdSe (wurzite) along the A-$\Gamma$-A path and the effect of an avoided crossing on the CDO.
    a) Band structure along the A-$\Gamma$-A path.
       The different markers depict the contribution of different WFs for three selected bands.
       Their size is proportional to the corresponding overlap.
    The gray rectangle (extended by a factor of 20 in each direction) indicates the region of the avoided crossing presented in b).
    c) Two selected matrix elements of the CDO in Wannier gauge.
        The gray stripe (extended by a factor of 20) indicates the region of the avoided crossing presented in d).
    e) Absolute values of the density matrix in Hamiltonian gauge and f) the dephasing operator in Wannier gauge for the $\bk$-points $\mathrm{P}$ and $\mathrm{Q}$ indicated in d).
    c-f) are based on a simulation of a pulse with $E_0 = 0.9\,\mathrm{V}/\mathrm{nm}$ at time $22.8\,\mathrm{fs}$ for a constant dephasing of $T_2=10\,\mathrm{fs}$.}
    \label{fig:BandstructureDephasing}
\end{figure}

\section{Implementation details \label{sec:implementationDetails} }
We denote the number of $\mathbf{k}$-points used during the propagation as $N^{\mathbf{k}} \equiv N^{\mathbf{k}}_1\times N^{\mathbf{k}}_2 \times N^{\mathbf{k}}_3$.
For both bases, we start our calculation at zero temperature, i.e., fully occupied valence and empty conduction bands.
The time derivative is calculated numerically and integrated using a Runge-Kutta (RK) 4/5 scheme with adaptive step size \cite{fehlberg_klassische_1970}.
We apply the dephasing operator during every time step of our numerical integration.
This accounts for possible rapid changes of the dephasing operator, which we discuss in Sec.~\ref{sec:dephasingOperator}.
The consistent inclusion of the dephasing combined with the adaptive step size solver allows for a uniform treatment of a $\bk$-dependent dephasing time $T_2(\bk)$, which may span multiple orders of magnitude in the BZ \cite{korolev_unveiling_2024,langer_lightwave-driven_2016}. 

We denote the time-dependent occupations of the $i$-th WF as $n^{\mathrm{W}\mathbf{k}}_i = \rho^{\mathbf{k}}_{ii}$ and track the Hamiltonian band-occupations, which are the diagonal elements of the density matrix in the Hamiltonian gauge, i.e., $n^{\mathrm{H}\mathbf{k}}_i = \rho^{\mathrm{H}\mathbf{k}}_{ii}$.

\subsection{Stationary basis}
In the SB, we evaluate the Hamiltonian and dipole matrix elements as well as the $U^\bk$ beforehand on a fixed MP grid.
The time derivative of the density matrix is calculated separately for the coherent and incoherent parts.
As we consider only linearly polarized light pulses here, the gradient of the density matrix in Eq.~\eqref{eq:SBE_fixed} is approximated by a one-dimensional five-point stencil.
Equation~\eqref{eq:SBE_fixed} decouples in momentum space orthogonal to the polarization direction, and was therefore parallelized. 
Due to the matrix multiplications involved, the computational complexity to calculate the derivative including dephasing is $\mathcal{O}(N^\mathbf{k} n_\mathrm{W}^3)$.
We note that for general pulses, e.g., with elliptical polarization, an efficient parallelization and the $\bk$-gradient calculation are more involved than described here. 

\subsection{Comoving basis}
The time derivative of the density matrix in the CB is calculated in three separate steps.
First, we calculate all the required Hamiltonian and dipole matrix elements, according to Eqs.~(\ref{eq:WanH}--\ref{eq:WandHdk}) for the crystal momentum $\mathbf{k} + \mathbf{A}$.
As a direct evaluation of the Fourier transform for all $(\bk + \mathbf{A})$-points is computationally inefficient, we calculate all desired matrix elements via FFTs.
To do so, we divide the MP grid into uniform subgrids with a suitable dimension to include all non-zero position space Wannier matrix elements within the Wigner-Seitz super cell.
The $\bA$-shifted expectation values are obtained by multiplying the real-space matrix elements with appropriate phase factors before applying the FFT on the subgrid.
For instance, the Hamiltonian matrix elements in the Wannier gauge, see Eq.~\eqref{eq:WanH}, are calculated for a subgrid as
\begin{equation}
    H^{\mathrm{W}\mathbf{k}+\mathbf{s} + \mathbf{A}(t)}_{mn} =
        \mathrm{FFT}_{\mathbf{R}}\left(\braW Rm \hat{H}_0 \ketW 0n \exmi{(\mathbf{s}+\mathbf{A})\mathbf{R}} \right)(\bk)
,
\end{equation}
where $\mathrm{FFT}_{\mathbf{R}}(\cdot)(\bk)$ denotes the FFT from position space to momentum space, and $\mathbf{s}$ is the shift of that subgrid with respect to the $\Gamma$-point.
Secondly, we evaluate the coherent part of the time derivative defined by Eq.~\eqref{eq:movingFrameVonNeumann}.
Thirdly, we incorporate the dephasing as described in Sec.~\ref{sec:OrigDephasingOperator}.
The transformation matrices $U^{\bk+\mathbf{A}}$, see Eq.~\eqref{eq:rhoWtoH}, are obtained by diagonalization of the Hamiltonian $H^{\bk+\mathbf{A}}$.
This results in a computational complexity of $\mathcal{O}(n_{\mathrm{W}}^3)$ to apply the dephasing at a single $\mathbf{k}$-point.
The overall computational complexity to calculate the time derivative of the density matrix is therefore
$\mathcal{O}\left( N^{\mathbf{k}} \left[n_\mathrm{W}^3    + n_\mathrm{W}^2 \log N_\mathrm{W}\right] \right)$,
where $N_\mathrm{W}$ is the number of grid points of the FFT.
We note that in typical SBE simulations, the second term is smaller than the first one, leading to the same computational complexity as in the SB.
In our implementation \cite{thuemmler_2025}, we parallelized the numerical integration of the density matrix over the subgrids.

\section{Dephasing in detail} \label{sec:dephasingOperator}
The Wannierization procedure \cite{marzari_1997} solves the long-standing issue of calculating a smooth and periodic structure gauge, required to obtain converged high-harmonic spectra \cite{yue_2020}.
However, the dephasing operator is still subject to the original band structure in the Hamiltonian gauge.
This leads to an ill-defined CDO at $\bk$-points with degenerate energies and will cause numerical difficulties when two or more bands come energetically close to each other.

\subsection{Ill-defined dephasing at points with degenerate energies}
We demonstrate the effect of the improper definition of the CDO, see Eq.~\eqref{eq:origDephasing}, for degenerate bands by an example.
These degeneracies will arise especially at symmetry points of the crystal.
We consider a diagonalized Hamiltonian with two degenerate eigenvalues, i.e.,
\begin{align}
    H^{\mathrm{H}} = U H^\mathrm{W} U^\dagger = \mathrm{diag}(E, E, E_2, E_3, \dots)
.
\end{align}
In the degenerate subspace, we can apply a rotation
\begin{align}
    U_r = \begin{pmatrix}  a & -b & 0 \\ b & a & 0 \\ 0 & 0 & \mathbb{1}  \end{pmatrix}
    \equiv \begin{pmatrix} \cos \alpha & -\sin \alpha & 0 \\ \sin \alpha & \cos \alpha & 0 \\ 0 & 0 & \mathbb{1}  \end{pmatrix}
.
\end{align}
so that $\tilde{U} = U_rU$ still diagonalizes $H^\mathrm{W}$. 
In the following, we only consider the non-trivial $2\times 2$ sub-block of $U_r$, where we study the effect of the rotation on the dephasing operator at an exemplary density matrix
\begin{align}
    \rho^{\mathrm{W}} = \begin{pmatrix} \rho_{11} & 0 \\ 0 & \rho_{22} \end{pmatrix}
.
\end{align}
Transforming it into the Hamiltonian gauge, we obtain
\begin{align}
    \rho^{\mathrm{H}} = \begin{pmatrix} a^2\rho_{11} + b^2 \rho_{22} & ab(\rho_{11} - \rho_{22}) \\ 
                                        ab(\rho_{11} - \rho_{22}) & b^2 \rho_{11} + a^2\rho_{22} \end{pmatrix}
,
\end{align}
and thus the dephasing reads
\begin{align}
    \left(\partial_t \rho^{\mathrm{H}} \right)_{\mathrm{Deph}} = \frac{ab(-\rho_{11} + \rho_{22})}{T_2}
     \begin{pmatrix} 0 & 1 \\ 
                     1 & 0 \end{pmatrix}
.
\end{align}
In Wannier gauge, after we substitute $a$ and $b$ for the trigonometric functions,
\begin{align}
    \left(\partial_t\rho^{\mathrm{W}} \right)_{\mathrm{Deph}} = \frac{\rho_{22} - \rho_{11} }{ 4 T_2}
     \begin{pmatrix} 2\sin^2(2\alpha) & \sin(4\alpha)  \\ 
                      \sin(4\alpha)   & -2\sin^2(2\alpha)  \end{pmatrix}
,
\end{align}
we see that the result strongly depends on $\alpha$.
In this simplified example, the subspace (described by $\alpha$) is not determined, leading to an ambiguity of the dephasing operator.
When considering the full band structure, it might be possible to fix the degenerate eigenspaces similar to what is done in degenerate perturbation theory \cite{klein_degenerate_1974}, but this is computationally expensive.
Additionally, for the CB it must be done at every time step, and its hard to ensure its reliability automatically.
\begin{figure*}[htbp!]
    \centering
    \includegraphics[width=0.99\textwidth]{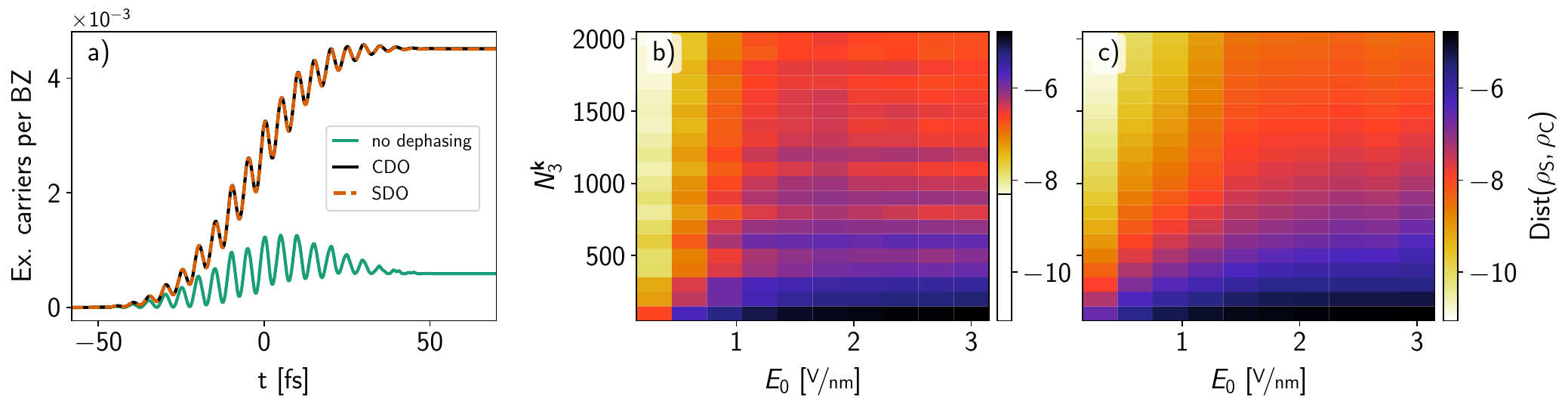}
    \caption{a) Time evolution of the excited carriers on a $30\times 30\times 100$ MP grid and $E_0 = 0.9\,\mathrm{V/nm}$ for the different dephasing operators.
    b, c) Distance of the density matrix in the SB to the reference solution in the CB for varying $N^{\mathbf{k}}_3$ and field strength $E_0$ for
    b) the CDO and
    c) the SDO.
    }
    \label{fig:EnKStability}
\end{figure*}

\subsection{Rapid change of dephasing close to avoided crossings}
\label{sec:rapidChangeDephasing}
In case of energetically close bands, the CDO is undergoing rapid changes.
We chose CdSe, which is subject of recent HHG experiments \cite{nakagawa_size-controlled_2022,gopalakrishna_tracing_2023}, as an example as it features an avoided crossing along the \agap, close to $\Gamma$.
We asserted that the avoided crossing is not an artifact of the Wannierization but is also present in the DFT-calculation.
The details of the parameterization are given in Sec.~\ref{sec:numericalResults}.
Within a tiny region of the BZ around the avoided crossing, see Fig.~\ref{fig:BandstructureDephasing}b), the composition of the band structure in terms of WFs is fully rearranged from WFs two and four to WFs two and three.
As the density matrix varies slowly in the Wannier gauge, this abrupt change manifests in a significant occupation of the non-diagonal elements of the density matrix in Hamiltonian gauge for the involved bands.
We illustrate this situation in Fig.~\ref{fig:BandstructureDephasing}e). 
Right before the avoided crossing (P) the density matrix in Hamiltonian gauge mainly shows polarizations between the valence and the conduction bands, whereas at the avoided crossing (Q), bands two, three, and four are strongly mixed.
As the CDO is proportional to the non-diagonal elements in the Hamiltonian gauge, this results in an enlargement of the dephasing in Wannier gauge, see Fig.~\ref{fig:BandstructureDephasing}f).
Compared to the $\bk$-dependence of the dephasing term of the transition between the second and seventh WFs (orange line of Fig.~\ref{fig:BandstructureDephasing}), which varies moderately across the whole \agap, the dephasing in the avoided crossing (black line) is extremely localized.
We used $\Nkc=10,000$ to resolve this peak which extends over approximately 0.5\% of the \agap.
Although this rapid change of the CDO is not erroneous, but rather a consequence of the avoided crossing, it will drastically decrease the stability of the numerical integration and lead to an increase of the computation time, while not having a significant effect on the macroscopic current, as we will see in Sec.~\ref{sec:numericalResults}.

\subsection{Soothed dephasing}
In the last section, we discussed why the CDO is ill-defined in the presence of band crossings and its rapid changes in the vicinity of avoided crossings.
Both situations suggest to suppress the dephasing in the case of energetically close bands.
As we want to modify the dephasing operator as little as possible and still require its smoothness in momentum space, we propose a Gaussian-like soothed dephasing operator (SDO)
\begin{equation}
    \label{eq:wellDefinedDephasing}
    \left( \partial_t {\rho^{\mathrm{H}\mathbf{k}}_{mn}} \right)_{\mathrm{Deph}} = - \frac{1 - \exp\left\{ - \left(\frac{E^{\mathbf{k}}_m-E^{\mathbf{k}}_n}{w_{\mathrm{S}}}\right)^2 \right\} }{T_2}  \rho^{\mathrm{H}\mathbf{k}}_{mn}
\end{equation}
with a soothing width $w_{\mathrm{S}}$, i.e., we reduce the dephasing between two states only, when the corresponding eigenenergies are close to each other.
This definition also ensures that only non-diagonal elements of the density matrix are damped.
It was reasoned before that the dephasing time is likely to be energy dependent \cite{fischetti_monte_1988,vu_light-induced_2004} and, in order to define a electromagnetically gauge independent dephasing, it was scaled with the energy difference between the involved bands \cite{wismer_gauge-independent_2018}.
However, our approach does not change the dephasing mechanism, but rather improves the widely used CDO, without significantly affecting the dynamics.

The band degeneracies pose the analogous problem of ambiguously defined occupation numbers.
To make them uniquely defined and consistent with the modification of the dephasing operator, we propose Gaussian-like weighting of the occupation numbers
\begin{equation} \label{eq:redefBandOcc}
    \bar{n}^{\mathrm{H}\mathbf{k}}_i = \frac{\sum\limits_{j} n^{\mathrm{H}\mathbf{k}}_j \exp\left\{ -\left( \frac{E^{\mathbf{k}}_i-E^{\mathbf{k}}_j}{w_\mathrm{S}}\right)^2 \right\} }
                     {\sum\limits_{j} \exp\left\{ -\left(\frac{E^{\mathbf{k}}_i-E^{\mathbf{k}}_j}{w_\mathrm{S}}\right)^2 \right\} }
\end{equation}
with the soothing width $w_\mathrm{S}$.
In \ref{app:Occupations} we demonstrate that Eq.~\eqref{eq:redefBandOcc} removes unreasonable oscillations from the occupation numbers in the Hamiltonian gauge.

\section{Numerical results \label{sec:numericalResults} }
We performed density functional theory (DFT) calculations for CdSe without spin-orbit coupling on a $9\,\times9\times9$ MP grid using Quantum Espresso 7.2 \cite{giannozzi_2020} based on norm-conserving pseudo-potentials \cite{hamann_2013,schlipf_2015} using a density cut-off of 50\,Ry. \
Using these calculations, we separately Wannierized the six highest valence and the two lowest conduction bands (after disentanglement \cite{souza_2001}) using Wannier90 \cite{marzari_2012,pizzi_2020} and shifted the band gap to its experimental value of $1.75\,\mathrm{eV}$ \cite{ninomiya_optical_1995}.
We again refer to the interpolated band structure shown in Fig.~\ref{fig:BandstructureDephasing}.
To concentrate the discussion on the avoided crossing, see Fig.~\ref{fig:BandstructureDephasing}, we orient the driving laser polarization along the \agap, i.e., the c-axis of the crystal (sampled by $\Nkc$ points).
We note that close to the high-symmetry point $A$ a similar rapid change in the band composition occurs, which will influence the convergence of the calculations with respect to the driving field strength.
The numerical artifacts arising at this crossing are less pronounced, as it is far from the $\Gamma$-point, where the carrier concentration is always much weaker.

We used a pulse of with a $\cos^2$ envelope for the vector potential
\begin{align}
    \bA(t) = \frac{E_0}{\omega} \cos^2\left(\frac{\pi t}{2 \pulsewidth}\right) \sin(\omega t) \hat{\mathbf{e}}_3
    ,
\end{align}
with a full-width half-maximum $\pulsewidth= 6\times\nicefrac{2\pi}{\omega}$ of six optical cycles, where $\omega$ is the fundamental frequency and $E_0$ the peak intensity.
For our chosen wavelength of $\lambda = 3\,\mu\mathrm{m}$, this corresponds to a full-width half-maximum of $\pulsewidth\approx 60.0\,\mathrm{fs}$.
In our simulations, we always use a dephasing time of $T_2 = 10\,\mathrm{fs}$, which is comparatively larger than in recent studies \cite{vampa_2014,silva_2019,heide_probing_2022}, but already showcases the problems of the CDO.
We confirmed that shorter dephasing times lead to even more pronounced numerical issues. 
The soothing width was set to $w_\mathrm{S}=25\,\mathrm{meV}$.
The qualitative behavior of the SDO results did not change for different choices of $w_{\mathrm{S}}$ in the range $10\,\mathrm{meV} \leq w_{\mathrm{S}} \leq 100\,\mathrm{meV}$.
During the numerical integration of the density matrix, we employ a relative and absolute error threshold for the RK 4/5 scheme of $10^{-10}$.
We have found that all numerical problems persist even with smaller integration thresholds.
In the following, $\rfix$ and $\rmove$ are used to indicate the numerically calculated density matrix in the SB via Eq.~\eqref{eq:SBEfixedCoherent} and the CB via Eq. \eqref{eq:SBEMovingCurrent}, respectively.

\subsection{Convergence of the calculations} \label{sec:StabilityInTime}
Having established all necessary methods and quantities, we now turn to the stability of the numerical simulations.
Figure \ref{fig:EnKStability}a) depicts the number of excited carriers along the \agap{} computed on a MP grid of $30\times30\times100$ in the CB.
We see typical Rabi oscillations for all three dephasing configurations, i.e., without dephasing, with the CDO as well as for the SDO.
As expected, the inclusion of dephasing leads to higher populations, as the Rabi oscillations are damped and carriers build up in the conduction bands.
Most importantly, there is no visible difference between the two dephasing approaches.
We elaborate on the changes induced by the soothing of the dephasing in Sec.~\ref{sec:comparisonDephasing}.

To assess the stability of the propagation of the density matrix in the SB, we compare it to the one in the CB.
As both approaches are only directly comparable at times with $\bA=\mathbf{0}$, we evaluate their element-wise difference at the $N_t = 25$ time steps for which this condition holds for our pulse.
To make a fair comparison between the different values of $\Nkc$, we evaluate this distance metric always on the same 100 $\bk$-points.
The density matrix of the CB serves as a reference, as the different $\mathbf{k}$-points are decoupled, and an increase of $N^{\bk}$ will not affect the accuracy of a single point.
Because the integration uses an adaptive step size, its numerical error is bounded by the predefined error threshold.
We perform our comparison using the distance metric
\begin{equation}
\label{eq:DistDensity}
    \mathrm{Dist}(\rfix, \rmove)= \frac{ \sqrt{
    \sum\limits_{\bk,t_i} ||\rfix(t_i) - \rmove(t_i)||^2 } }
    { N^{\bk} n_{\mathrm{W}}^2 N_t}
    %{\sum\limits_{k} ||\rho^{\mathrm{W}}_{\mathrm{H}, \mathbf{k}}(t)||^2}}
    ,
\end{equation}
where $||\cdot||$ denotes the Frobenius norm.
This distance describes the average deviation per matrix element of the two density matrices.
It is invariant under basis transforms, e.g., going from the Wannier to the Bloch gauge.

Figure~\ref{fig:EnKStability} depicts the time-averaged distances for different $\Nkc$ and different maximum field strengths.
While the error in the case of the CDO, see panel b), generally increases with increasing field strength and decreases with increasing $N^{\bk}_3$, it does not converge to the numerical accuracy, even for small field strengths.
We marked the error limit of roughly $10^{-8}$ in the scale of the colorbar.
Some values for $N^{\bk}_3$ and $E_0$ clearly result in a decreased numerical stability.
For instance, $E_0=1.8\,\mathrm{V/nm}$ leads to a worse convergence than $E_0=1.5\,\mathrm{V/nm}$ and $E_0=2.1\,\mathrm{V/nm}$.
This can be understood as follows: 
The carriers are predominantly excited near the $\Gamma$-point and accelerated toward the $A$-point where they pass over the already discussed avoided crossing for all investigated field strengths. 
For $E_0=1.8\,\mathrm{V/nm}$, these carriers will almost exactly reach the $A$-point, where another energetically close subspace is located. 
Since this happens when the electric field is maximal, the carriers spend a relatively large amount of time at this point with an ill-defined dephasing operator and hence the numerical accuracy is decreased.
The convergence with respect to $N^{\bk}_3$ is non-monotonic, too, e.g. the error for $N^{\bk}_3=800$ is significantly lower than for $N^{\bk}_3=900$ as by chance the CDO is especially badly resolved for the (avoided) crossing.
We underline that the outliers in the convergence are purely determined by the positions of the (avoided) crossings.
In comparison, the SDO, see panel c), exhibits a well-defined convergence behavior.
An increase in field strength can now be compensated for by an increase of $N^{\bk}_3$ to reach the desired numerical error threshold.
The overall agreement between stationary and CB is around two orders of magnitude better compared to the CDO.
Tracing the error over time shows the partially erratic behavior of the CDO as well as the limited convergence even for a single field strength. 
We included an analysis of the time-resolved evolution of the distance in \ref{app:StabilityInTime}.
\begin{figure}[tb]
    \centering
    \includegraphics[width=\columnwidth]{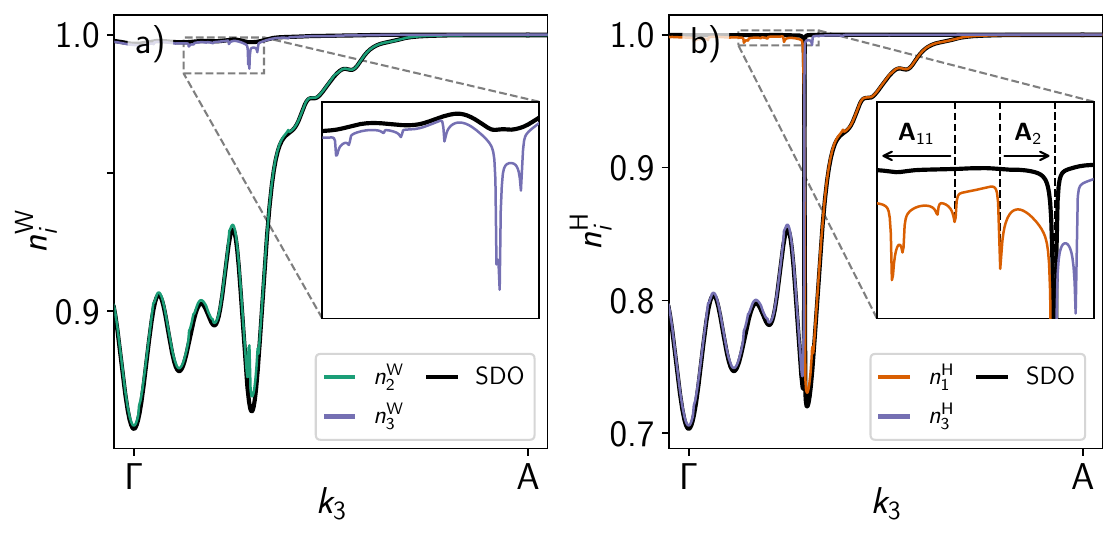}
    \caption{Selected occupations in a) Wannier and b) Hamiltonian gauge at the end of a pulse with $E_0=0.9\,\mathrm{V/nm}$. 
    The density matrix was propagated in the CB along the \agap.
    The colored and black lines were calculated in the CDO and the SDO, respectively.
    The WFs which correspond to the shown occupations participate in the avoided crossing.
    }
    \label{fig:momentumStability}
\end{figure}

\subsection{Occupation numbers}
To shed more light on the numerical stability, we now consider the $\bk$-dependence of the occupation numbers in the CB to demonstrate the importance of the SDO.
Figure~\ref{fig:momentumStability} shows the occupations in Wannier and Hamiltonian gauge at the end of the pulse.
Generally, they vary slowly throughout the \agap, but they also feature multiple kinks, with the most prominent one at the position of the band crossing.
This position coincides with the peak of the CDO, discussed in Sec.~\ref{sec:rapidChangeDephasing}. 
Most of the time, the carriers pass over this region quickly, but twice during every cycle when the amplitude of the vector potential is maximal and the electric field vanishes, the carriers which are in the vicinity of the avoided crossing are dephased for a relatively long time.
This causes the formation of two kinks per cycle at exactly the distance between the position of the avoided crossing and the current extremum of the vector potential, see Eq.~\eqref{eq:sbcbTransform}.
For example, the kink in $n^{\mathrm{H}}_1$ which originates from the second extremum of the vector potential $\bA_2$ at $t \approx -51.9\,\mathrm{fs}$, see inset of Fig.~\ref{fig:momentumStability}b), is located in a BZ region where the occupations are close to one. 
This region is shifted directly into the avoided crossing region, where $n^{\mathrm{H}}_1$ drops significantly. 
Due to the strong dephasing, the Rabi oscillations are damped and the kink is formed.
For extrema of the vector potential of odd order, the same effect is observed but from the avoided crossing on the other side of the BZ, e.g., the kink arising during the eleventh extremum is indicated by $\bA_{11}$ ($t \approx -7.5\,\mathrm{fs}$).
When we compare the CDO (colored lines) with the SDO (black lines), we see that the kinks are suppressed and otherwise both curves coincide very well in both the Wannier and Hamiltonian gauge.
This shows that the modification of the dephasing operator preserves the main features of the time evolution of the density matrix.

We also performed the analysis for the SB, where numerical instabilities arise only for the CDO.
In \ref{app:StabilityInMomentumFixed}, we provide an example for 400 $\bk$-points along the \agap{}. 
The sharp kinks cannot be resolved, and instead we see fast oscillations of the occupation numbers that might even exceed one or go below zero.
They are the reason for the slow convergence shown in Fig.~\ref{fig:EnKStability}.
For the SDO, the fast oscillations vanish completely and the solutions coincide with those obtained from the calculations in the CB.
\begin{figure}[tb]
    \centering
    \includegraphics[width=\columnwidth]{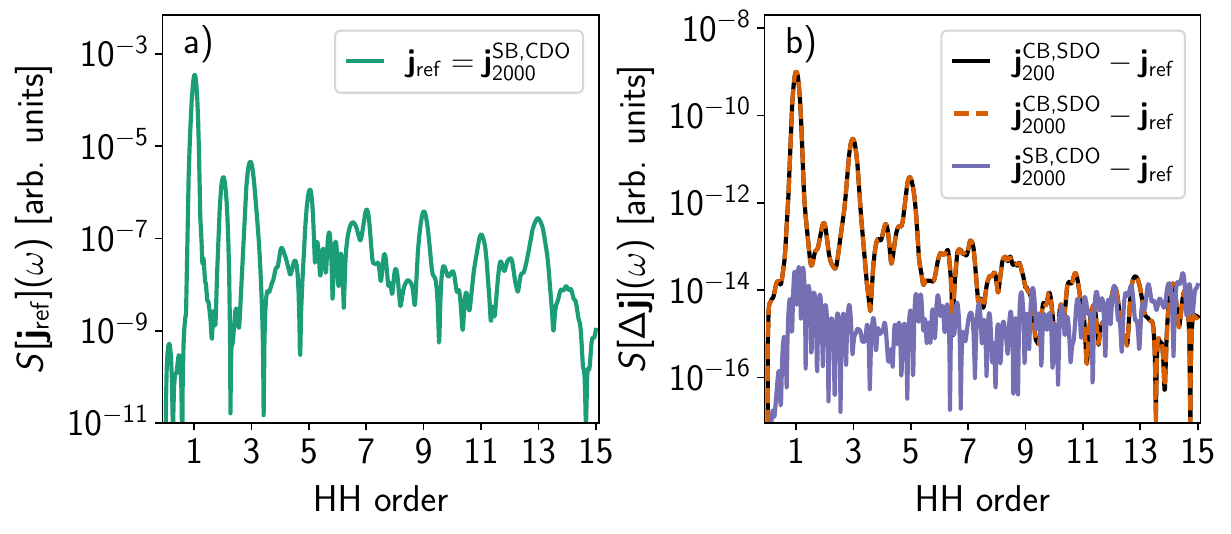}
    \caption{High-harmonic spectrum ($E_0=0.9\,\mathrm{V/nm}$) of CdSe a) for the most converged simulation in the CB for CDO and b) the difference to the simulation in the SB (purple line), as well as the simulation in SDO and either $N^{\bk}_3=200$ (black line) or $N^{\bk}_3=2000$ (orange dashed line).
    We note that the intensity of the difference spectra in b) are at least six orders of magnitude smaller than the intensity of the reference spectrum.
    }
    \label{fig:spectrum}
\end{figure}

\subsection{Impact on the high-harmonic spectrum} \label{sec:comparisonDephasing}
The kinks in the distribution of excited carriers which are created by the CDO seem implausible, but will hardly affect the total amount of excited carriers. 
However, the high-harmonic spectrum
\begin{align}
    \label{eq:spectrum}
    \left(S[\bj]\right)(\omega) \propto \omega^2 \big|\big|\bj(\omega)\big|\big|^{2}
\end{align}
might be sensitive to small perturbations as it is usually considered across multiple orders of magnitude.
In Fig.~\ref{fig:spectrum} we consider HHG spectra calculated for grid-sizes of $30\times30\times\Nkc$. 
The numerically converged spectrum $S[\bj_{\mathrm{ref}}]$ calculated in the CB with the CDO for $N_3^{\bk}=2000$ is displayed in panel a). 
As the spectra obtained for the SDO are very similar, we do not plot them directly, but instead we show the spectrum of the difference of the calculated currents in panel b), i.e., $S[\bj_{N_3^{\bk}}^{\mathrm{CB,SDO}} - \bj_{\mathrm{ref}}]$ for $N_3^{\bk}=200$ (black solid line) and $N_3^{\bk}=2000$ (orange dashed line).
The spectra practically coincide for the two different dephasing approaches, with a deviation of around $10^{-6}$, which is reflected in the different scale of panel b). 
This can be understood when we consider the typical separation of the spectrum into the below band gap part which is governed by the intraband current and the above band gap part created by the interband polarization \cite{golde_high_2008}. 
The latter is only weakly affected by both dephasing operators, as we chose a comparatively small soothing width $(w_\mathrm{S} =25\,\mathrm{meV})$. 
The intra-band current is only important for the first few harmonics and hence the high frequency kinks created by the CDO are heavily suppressed.
For comparison, the difference to the SB $S[\bj^{\mathrm{SB,CDO}}_{\Nkc}-\bj_{\mathrm{ref}}]$ calculated for $\Nkc = 2000$ is plotted as a purple line and limited by the numerical precision of our simulation.

\subsection{Computation time}
\begin{figure}[tb]
    \centering
    \includegraphics[width=\columnwidth]{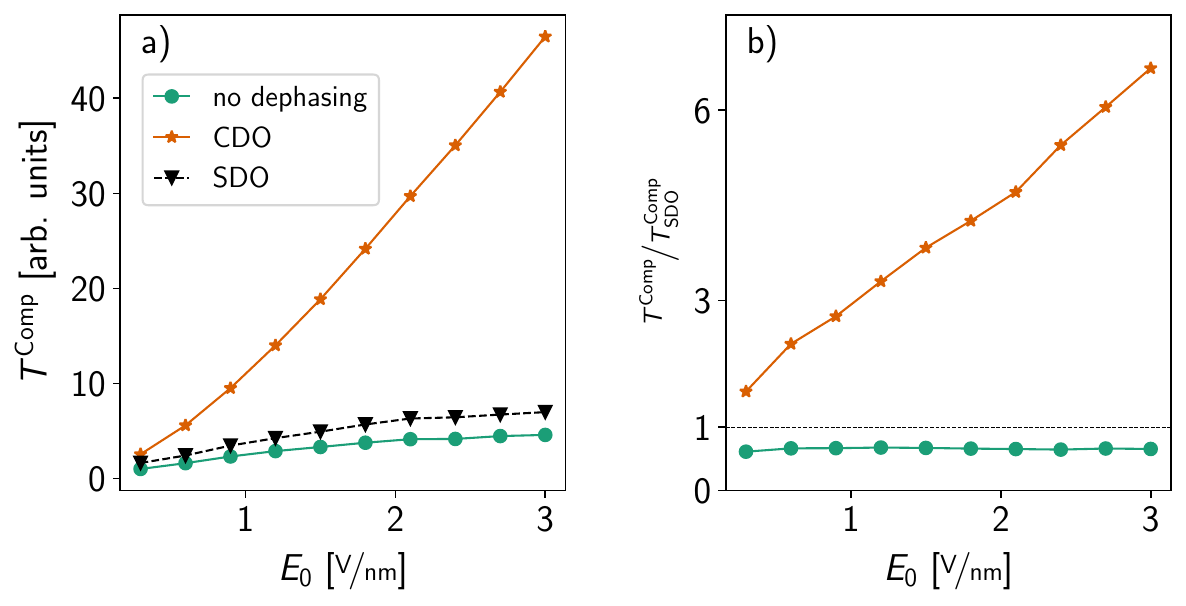}
    \caption{Computation times to simulate a single pulse in the CB ($N^{\bk}_3=100$) for all dephasing types and different maximum field strengths.
        a) Normalized to the overall lowest computation time.
        b) Normalized to the computation time of the propagation with the SDO.
        }
    \label{fig:timingComparison}
\end{figure}

The SDO does not only lead to numerically more stable results, but also reduces the computational effort immensely.
We consider only the computation times of the CB as the SB computation time is intertwined with its numerical instability.
In Fig.~\ref{fig:timingComparison} we present the relative calculation times for the simulation of one pulse for the three different dephasing types as function of the maximum field strength.
The inclusion of the CDO results in an overhead compared to the coherent propagation, which scales approximately linearly with the field strength.
Here, the numerical resolution of the CDO at the avoided crossings is responsible for this scaling behavior.
At these points, the rapid change of the CDO $\rho_{\mathrm{Deph}}^{\bk+\bA}$ is counteracted by a reduction of the adaptive step size.
To understand this quantitatively, consider the incoherent propagation in a super-operator notation $\partial_t \rho^{k} = L_{\mathrm{Deph}}(k +A)\rho^{k}$ in one dimension.
The error estimate of a RK 4/5 time step $\Delta t$ at time $t$ is given by 
$\mathcal{O}\left( (\Delta t)^5 |E(t)|^4L^{(4)}_{\mathrm{Deph}}(k + A(t)) \right)$,
where $L^{(4)}_{\mathrm{Deph}}$ denotes the fourth derivative of $L_{\mathrm{Deph}}(k)$ with respect to $k$ \cite{fehlberg_klassische_1970}.
To reach a predefined constant error, the solver must scale the time step as $E^{-0.8}(t) \propto E_0^{-0.8}$, resulting in a corresponding increase of the number of time steps required to propagate the density matrix in the CB through the avoided crossing.
The resolution of those crossings dominates the integration effort, as we propagate the density matrix in joined MP grids (to apply the FFT), where the step size must be reduced as soon as any $(\bk +\bA)$-point approaches an hard-to-integrate (avoided) crossing.
In our timing measurements, the computation time increases approximately linear in the field strength, which we attribute to the additionally increased occupations for higher field strengths.
When we use the SDO instead of the CDO, the original scaling behavior of the coherent propagation is restored, as the SDO does not exhibit these rapid changes.
Panel b) reveals that the inclusion of the dephasing via SDO leads only to a $\sim 50\%$ increased computation time compared to the coherent propagation.

\section{Conclusion \label{sec:sao}}
The Wannierization of the SBEs combined with a comoving basis (CB) is a reliable tool for the description of HHG in solids.
We employed FFTs to efficiently obtain the matrix elements during the numerical propagation in the CB, resulting in a computational cost similar to that of the stationary basis (SB).
We showed that the consistent inclusion of the commonly used constant dephasing operator (CDO) leads to a drastic increase in computation time in the CB, as well as to numerical instabilities in the SB. 
This behavior was traced back to a strong enhancement of the matrix elements of the CDO close to avoided crossings and manifests itself in the formation of kinks in the carrier distribution. 
We proposed a soothed dephasing operator (SDO) that removes the kinks and the numerical instabilities.
It also immensely reduces the simulation time while keeping the high-harmonic spectrum unchanged and additionally circumvents the ill-defined nature of the CDO at degenerate points.
In summary, the SDO is simple, mathematically well defined, and numerically more robust than the CDO, while its still preserves all its desired features.
Therefore, we suggest to integrate the SBEs in the Wannier gauge with the SDO using an adaptive solver in the CB.
We provide our implementation for the CB in \cite{thuemmler_2025}.

\section*{Funding}
{\sloppy
The project was funded by the Deutsche Forschungs\-gemeinschaft (DFG, German Research Foundation) -- Project-ID 398816777 -- SFB 1375, A1.\par
}

\appendix
\section{Time-resolved convergence of the density matrix in the SB} \label{app:StabilityInTime}
\begin{figure}[htpb]
    \centering
    \includegraphics[width=\columnwidth]{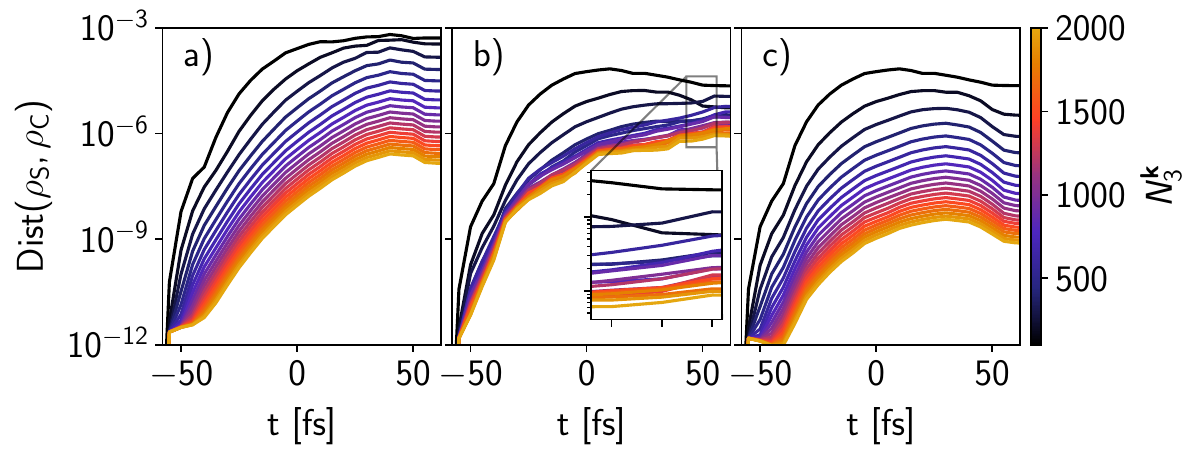}
    \caption{Time-resolved distance of the density matrix in the SB (varying $N^{\bk}_3$) to the density matrix in the CB ($N^{\bk}_3=100$) for $E_0=0.9\,\mathrm{V/nm}$.
    We either apply
        a) no dephasing, 
        b) the CDO or
        c) the SDO.
    }
    \label{fig:timeStability}
\end{figure}

In Fig.~\ref{fig:timeStability} a) to c), we show the time-resolved distances between the SB (varying $N^{\bk}_3$) and the CB ($N^{\bk}_3=100$) for the same 100 evenly placed $\mathbf{k}$-points along the \agap.
We included a simulation without dephasing in panel a) to underline that the convergence issues arise only due to the CDO, see panel b).
Although the overall error is still reduced for larger values $\Nkc$ when the CDO is included, the dependence on $N^{\bk}_3$ is not consistent over time.
It is not even converged to the reference solution for $\Nkc = 2000$.
The SDO, see panel c), leads to a significant improvement of the convergence behavior.
It restores the monotonicity of the convergence of the SB to the CB and the error is now close to the numerical threshold for the largest $N^{\bk}_3=2000$.

\section{Numerical instabilities of the density matrix in momentum space in the SB} \label{app:StabilityInMomentumFixed}
\begin{figure}[htpb]
    \centering
    \includegraphics[width=\columnwidth]{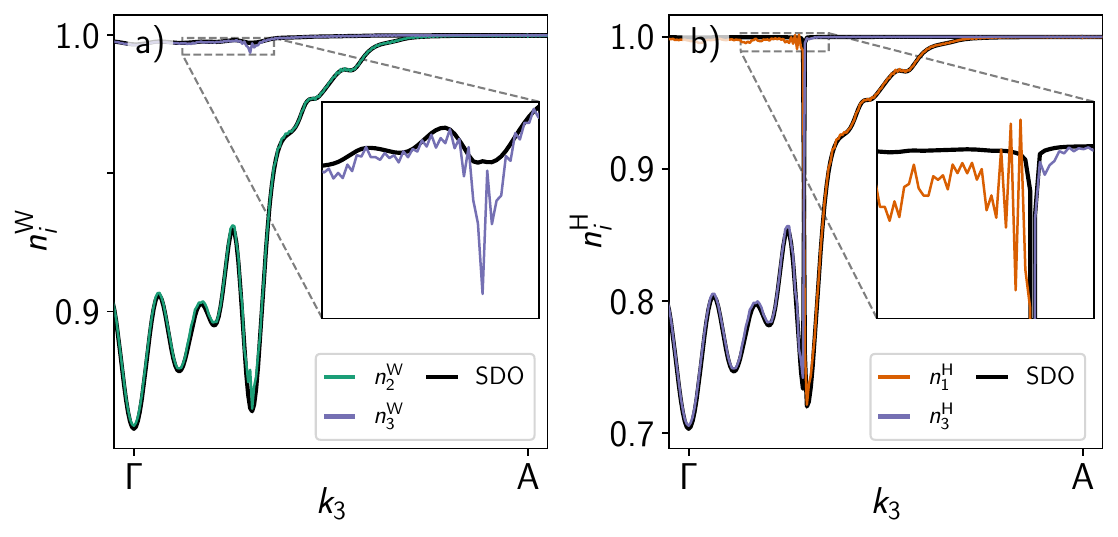}
    \caption{Selected occupations in a) the Wannier and b) the Hamiltonian gauge at the end of a pulse with $E_0=0.9\,\mathrm{V/nm}$ propagated in the SB along the \agap{} for $\Nkc =400$.
    The colored and black lines were calculated using the CDO and the SDO, respectively.}
    \label{fig:momentumStabilityFixed}
\end{figure}

Figure~\ref{fig:momentumStabilityFixed} depicts the same occupations at the end of the pulse as Fig.~\ref{fig:momentumStability}, except that the SB is employed instead of the CB during the simulation.
The CDO in the SB has severe numerical artifacts:
First, high frequency oscillations in $\bk$-space build up due to the discretization of the $\bk$-derivative.
Second, their amplitude is maximal around the kink positions of the CB, but spread throughout the full BZ.
Third, the deviation to the SDO approach is much larger.
And lastly, even non-physical occupations ($>1$) in the Hamiltonian gauge occur.
We have checked, that by increasing $\Nkc$ all these artifacts are reduced and a convergence to the CB, see Fig.~\ref{fig:momentumStability}, can be achieved.
Similar to the CB, there are none of these numerical problems with the SDO.

\section{Occupation number mixing in the Hamiltonian gauge} \label{app:Occupations}
\begin{figure}[htpb]
    \centering
    \includegraphics[width=\columnwidth]{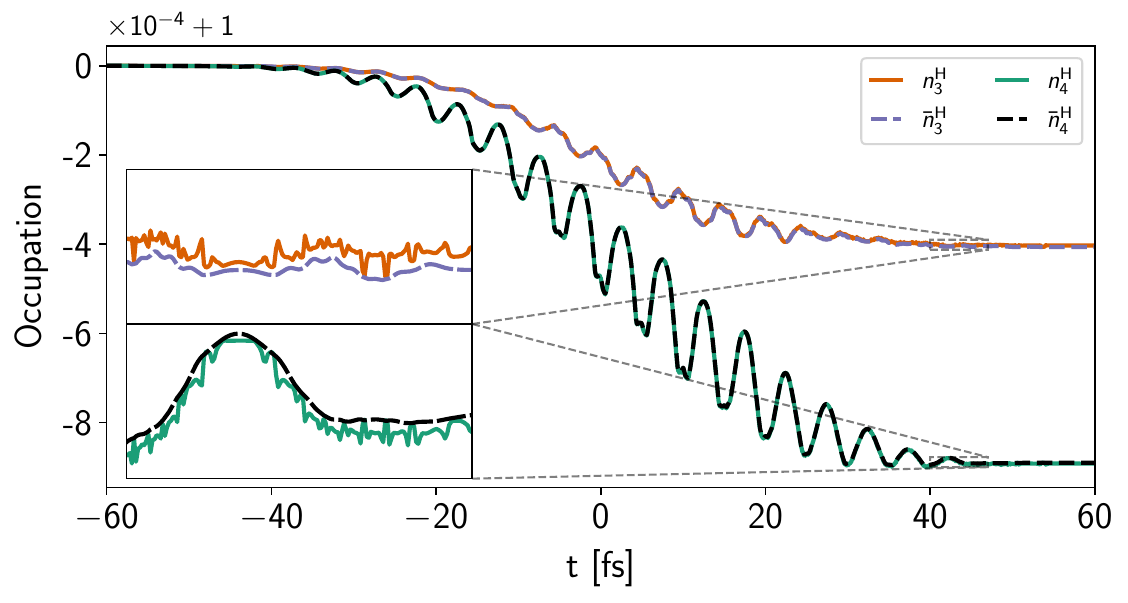}
    \caption{Averaged original (solid lines) and mixed occupation numbers (dashed lines) for two valence bands calculated on a $30\times 30\times 100$ MP grid. 
    The simulation was performed in the CB with the SDO and $E_0=0.9\,\mathrm{V}/\mathrm{nm}$.
    } 
    \label{fig:occupationNumbers}
\end{figure}

In Fig.~\ref{fig:occupationNumbers} we demonstrate the effect of Eq.~\eqref{eq:redefBandOcc} on the Hamiltonian band occupation numbers.
The bands for which the BZ-integrated occupations are depicted take part in the avoided crossing presented in Fig.~\ref{fig:BandstructureDephasing}.
Besides the typical oscillations, the integrated occupations exhibit additional high-frequency oscillations, whose sum cancels out.
These oscillations arise as the band occupations for a single $(\bk+\bA)$-point change rapidly while it passes an avoided crossing.
A resolution of these rapid changes would require much more than 1000 evenly placed $\bk$-points on the \agap{}, see Fig~\ref{fig:BandstructureDephasing}.
In comparison, the mixed occupation numbers show no high-frequency oscillations at all, as they are averaged in the vicinity of the (avoided) crossings.
The redefined occupation numbers are slightly shifted towards each due to their mixing.
We finally emphasize that in case of band degeneracies, the original occupation numbers are ill-defined.
In contrast, the redefinition leads to well-defined occupation numbers, since they are based only on the trace over the degenerate subspace, which is invariant under a change of basis.

\end{document}